\documentstyle[12pt]{article}
\def\TL{\hfil$\displaystyle{##}$}
\def\TR{$\displaystyle{{}##}$\hfil}

\def\comment#1{}
\def\fixit#1{}

\def\overleftrightarrow#1{\vbox{\ialign{##\crcr
     $\leftrightarrow$\crcr\noalign{\kern-0pt\nointerlineskip}
     $\hfil\displaystyle{#1}\hfil$\crcr}}}

\def\lsim{\mathrel{\mathstrut\smash{\ooalign{\raise2.5pt\hbox{$<$}\cr\lower2.5pt\hbox{$\sim$}}}}}
\def\gsim{\mathrel{\mathstrut\smash{\ooalign{\raise2.5pt\hbox{$>$}\cr\lower2.5pt\hbox{$\sim$}}}}}

\def\sqr#1#2{{\vcenter{\vbox{\hrule height.#2pt
         \hbox{\vrule width.#2pt height#1pt \kern#1pt
            \vrule width.#2pt}
         \hrule height.#2pt}}}}

\def\href#1#2{#2}  

\def\lbldef#1#2{\expandafter\gdef\csname #1\endcsname {#2}}
\def\eqn#1#2{\lbldef{#1}{(\ref{#1})}%
\begin{equation} #2 \label{#1} \end{equation}}
\def\eqalign#1{\vcenter{\openup1\jot
    \halign{\strut\span\TL & \span\TR\cr #1 \cr
   }}}

\def\comment#1{  \begin{raggedright}{\tt [#1]}\end{raggedright}}
\def\fixit#1{}

\def\comment#1{  \begin{raggedright}{\tt [#1]}\end{raggedright}}
\def\fixit#1{}

\begin{document}
\baselineskip=15.5pt
\pagestyle{plain}
\setcounter{page}{1}


\begin{titlepage}

\begin{flushright}
PUPT- 1983 \\  
hep-th/0104125
\end{flushright} 
\vfil

\begin{center}
{\Large Little String Thermodynamics}
\end{center}
\vspace{1.5cm}
\begin{center}
{\large Mukund Rangamani}
\end{center}
\begin{center}   
Joseph Henry Laboratories, Princeton University, Princeton,
NJ 08544.
\end{center}  
\vfil

\begin{center}
{\large Abstract}
\end{center}
\noindent
We argue that the holographic dual to little string theories 
at finite temperature suffers from a Gregory-Laflamme like 
instability, providing an alternative explanation to the 
results of hep-th/0012258. 
\vfil
\vfil
\vspace{6cm}
\begin{flushleft}
April 2001.
\end{flushleft}

\end{titlepage}
\newpage

\section{Introduction}
Little string theories (LSTs) are defined to be the 
theory on the world-volume of NS5-branes 
in the limit of vanishing string coupling $g_{str} \rightarrow 0$, 
\cite{seiberg} (for a review see \cite{aharony}).  In this limit 
it is argued that the world-volume degrees of freedom decouple from 
the bulk dynamics. The LSTs are non-local field theories \cite{kapustin}
without gravity and one hopes to understand the physical 
consequences of non-locality from them as a precursor to being able to 
address the question in the general setting of quantum gravity. 
An alternate definition of LSTs can be given by looking at 
string theories compactified on singular Calabi-Yau spaces \cite{gkp}. 

Despite having successfully defined decoupled non-local 
theories on the world-volume of NS5-branes, it has proven quite 
hard to extract substantial information about the dynamics of the theory.
Efforts in this direction include formulation of a light-cone description 
\cite{abs}, which has proved useful in extracting the chiral operator 
spectrum and the formulation of a holographic dual in terms of 
string theories in linear dilaton backgrounds \cite{abks}. The 
holographic description has been used to extract some correlation 
functions \cite{minsei}  and has been extended to more general theories 
with fewer supersymmetries \cite{gka, gkb}. 

In conventional field theories one learns much about the underlying 
physics by subjecting matter to extreme conditions, high 
energies or high temperatures. One would hope that under similar 
treatment one could learn something about the universal features of 
LSTs. In particular the thermal behavior of LSTs has been 
addressed by various authors. It was argued in the work of 
\cite{ms} that the LSTs at finite temperature can be realized in 
terms of the CGHS black hole \cite{cghs}. As a consequence of this 
the holographic prescription of \cite{abks} implies that the 
finite temperature LST is holographically dual to string theory on
the CHS tube \cite{chs} capped off by a horizon (the two-dimensional 
cigar geometry in Euclidean metric), along with a $SU(2)$ WZW model with 
level prescribed by the number of NS5-branes and a free CFT for the 
longitudinal directions of the brane. Actually, it is not quite 
true that this background describes the LST at finite temperature. 
In fact as we shall review LST in this background is precisely dual to 
LST at its Hagedorn temperature \cite{ms}. 
The tree level thermodynamics has therefore, entropy proportional to 
energy, implying an exponential growth of states\footnote{
The exponential growth of states of LST was noted in the DLCQ description
in \cite{ab}.}. To get a better understanding of the system one therefore 
has to look at higher order corrections to thermodynamics, which 
has been investigated in \cite{ho, br, ks,kkk}. We shall in the present 
work try to reinterpret some of the results presented in \cite{ks}. 

The punch line of \cite{ks} was that the thermal ensemble corresponding 
to LSTs at Hagedorn temperature exhibits an instability; it has 
negative specific heat. This was interpreted to signal the presence of 
a tachyonic mode in string theory about the background. Using the 
fact that the CFT in the background could be explicitly solved
(the Euclidean cigar background is a $SL(2)/U(1)$ coset), the authors 
exhibited a mode that wound once around the Euclidean time direction,
whose mass at tree level was shown to vanish \cite{ks, gkb}. 
The instability at one-loop was supposed to be due to this mode
becoming tachyonic (mass generation at one-loop is possible thanks to 
spacetime supersymmetry having been broken due 
to the thermal boundary conditions).

This is not a satisfactory state of affairs, since as we shall see
supergravity is perfectly valid at the energy scales under question.
It seems unnatural to have a geometry which is manifestly nice and 
smooth, wherein one would have a stringy tachyonic mode. 
We propose rather that the instability has its origins in 
supergravity itself. A concrete statement of the proposal is:

``The supergravity dual to LST at Hagedorn temperature suffers from a 
Gregory-Laflamme (GL) \cite{gla,glb}
like instability, thereby causing the thermal 
ensemble to be unstable.''

The GL instability in 
question is argued to be marginal, in the sense that it is a 
massless or zero frequency mode that one finds in the classical 
fluctuation analysis at tree level. We claim that this mode is potentially 
capable of acquiring a tachyonic mass at one-loop, taking 
responsibility for destabilizing the thermal ensemble. So we differ 
from the interpretation of \cite{ks} in origins of the massless mode; 
while the authors of \cite{ks} would 
like the massless mode to be coming from a stringy excitation which is 
localized at the tip of the Euclidean cigar, our contention is that 
it is just a metric perturbation about the classical background.

We have not managed to explicitly demonstrate that the metric fluctuations 
about the background of interest have indeed a zero frequency mode that 
is capable of explaining the origins of the instability. However, we 
argue based on a recent paper of Reall \cite{reall}; wherein it is 
claimed that near-extremal NS5-branes suffer from the GL instability, that is 
very plausible that in the decoupling limit for finite 
temperature LSTSs \cite{ms}, the relevant mode survives. Note that this 
interpretation is along the lines of the conjecture of 
\cite{gma, gmb}, arguing that thermodynamic instabilities in field 
theories should correspond to classical instabilities of the dual 
spacetime geometry.  

The plan of the paper is as follows: We first begin with a review 
of tree-level thermodynamics of LSTs and then proceed to describe 
the calculations of \cite{ks} briefly. We discuss their interpretation of 
the result and argue that the instability would persist if one 
took the supergravity limit of the string theory partition function 
derived in \cite{ks}. Hence the origins of the 
instability must lie within reach of supergravity
 and proceed to argue that this is 
indeed the case aided by the supporting evidence of \cite{reall}. 

\section{Review of LST thermodynamics}
\subsection{Classical thermodynamics}
The metric of the non-extremal NS5-brane in the string frame is \cite{hs}

\eqn{undstrmet}{\eqalign{
ds_{str} ^2 & = -f(r) dt^2 + dy_5^2 + A(r) \left( {dr^2 \over f(r)} +
 r^2 d\Omega_3^2
\right) \cr 
e^{2 \Phi} &= g_{str}^2 A(r) \cr
f(r) &= \left( 1 - {r_0^2 \over r^2} \right) \cr
A(r) &= \left( 1 + {N l_s^2\over r^2} \right)
}}

\noindent
The background has in addition a constant flux of the NS-NS field 
strength piercing the ${\bf S^3}$, which we shall not write 
explicitly. The parameter $r_0$ is the radius of 
the horizon and is related to the energy density above 
extremality as 

\eqn{energyden}{
{E \over V_5} = { 1 \over (2 \pi)^5 \alpha'^3} \left({ N \over g_{str}^2 }+ 
{r_0^2 \over g_{str}^2 \alpha'} \right).
}

The decoupling limit \cite{ms} is defined  
by scaling the asymptotic value of the string coupling and the horizon 
radius to zero, while holding the the energy density above extremality 
fixed in string units, {\it i.e.}, 

\eqn{decolim}{
g_{str} \rightarrow 0; \;\;\;\; r_0 \rightarrow 0 
; \;\;\;\;\; \mu = {r_0^2 \over g_{str}^2 l_s^2} \rightarrow \rm{fixed}.
}

To analyze this limit it is convenient to introduce a new coordinate
$r = r_0 \cosh \sigma $ (so that we only focus on the region of 
the spacetime exterior to the horizon), obtaining

\eqn{scalim}{\eqalign{
ds_{str}^2 &= -\tanh^2 \sigma  \;dt^2 + Nl_s^2 \; d\sigma^2 + Nl_s^2 \;
d\Omega_3^2 + dy_5^2 \cr
 e^{2\Phi} &= {N \over \mu \cosh^2 \sigma } 
}}

\noindent
Thus, as is clear from the metric in \scalim, the spacetime is the
direct product of the two-dimensional dilaton black hole \cite{ cghs}, 
a three-sphere of constant radius proportional to
$\sqrt{N \alpha'} $ and five dimensional flat space. 
The geometry is perfectly smooth, curvatures being small
everywhere (when the number of NS5-branes $N$ is taken to be large)
and simultaneously string perturbation theory is also 
good (so long as $\mu \gg N \gg 1$, since the dilaton is 
bounded from above by its value at the horizon, ${ N \over \mu}$).
The above geometry is supposed to be holographically dual to 
LST at finite temperature \cite{seiberg, abs, ms}. It is also 
closely related to the description of double scaled LST (DSLST) 
described in \cite{gka, gkb}.  

Continuing to Euclidean time, we obtain the the two dimensional cigar,
which asymptotes to the linear dilaton vacuum times a Euclidean timelike 
circle. This along with the rest of the spacetime can be described by an 
exact conformal field theory, with target space

\eqn{cft}{
H_3^+/U(1) \times SU(2)_N \times {\bf R}^5, 
}

\noindent
where $H_3^+ = SL(2,C)_N/SU(2)_N$, is the Euclidean cigar. 
Knowing the explicit solution to the metric we can read off as usual the 
temperature of the black hole by demanding the absence of a conical 
singularity at the tip $(\sigma = 0)$ in Euclidean time. This gives

\eqn{temp}{
\beta_H = { 1 \over T_H } = 2 \pi \sqrt{N \alpha'}.
}

The Hawking temperature of the black hole is thus calculated to be 
independent of the energy, implying degeneracy of the thermal 
ensemble. Note that this temperature is exactly the Hagedorn temperature 
of the LST. The fact that one can tune the energy and the temperature 
of the system independently indicates the proportionality of the entropy 
to the energy, implying that the free energy vanishes. The equation of 
state is the Hagedorn form
\eqn{eos}{
S = \beta_H E 
}

\noindent
which leads to an exponentially growing density of states, 
$\rho(E) \sim e^{\beta_H E}$. 
In order to understand the thermal ensemble one has to study 
quantum corrections to the above. Calculation of the one-loop 
free energy would tell us whether or not the thermal 
ensemble is stable. This calculation was carried out in a recent 
work of Kutasov and Sahakyan \cite{ks}, which we review in the next
section.

Since the background could be understood as a CFT, we can ask how is it 
that in the CFT we see that the equation of state is as given in \eos. 
It was argued in \cite{ks} that world-sheet supersymmetry of the 
CFT is sufficient to guarantee vanishing of the tree-level free 
energy\footnote{It is not sufficient in linear dilatonic backgrounds to 
claim that
the tree level partition function vanishes because of the infinite 
volume associated with the $SL(2,C)$ conformal Killing group of the sphere, 
for the volume of the cigar has a similar divergence, see \cite{ks} for 
details.}.

\subsection{One loop analysis} 
We would like to estimate the one-loop free energy of the gravitational 
background \scalim. One way to go about this would be to do a one loop 
analog of the Gibbons-Hawking \cite{gibhawk} calculation of evaluating 
the Euclidean action about the solution. But this is fraught with the 
usual problems of regulating the divergences in supergravity.
Nonetheless, we are aware of the fact that string theory provides a 
good regulator mechanism and one would hope that the 
string partition function evaluated in the background \scalim\ would 
contain valuable information about the thermal ensemble. 

Loop/stringy corrections to the Hagedorn density of states of 
LST were studied in \cite{ho,br, ks}. The basic argument was that 
finite energy corrections to the Hagedorn density of states,
$\rho(E) \sim e^{\beta_H E}$, would be of the form

\eqn{dos}{
\rho(E ) \sim E^{\alpha} e^{\beta_H E} \left( 1 + {\cal{O}}
\left({1 \over E}\right) \right).
}

\noindent
The primary question is ``What is the sign of $\alpha$?''
A negative value of $\alpha$ would indicate that the thermal ensemble is 
unstable (since the specific heat would be negative). Indeed 
the explicit calculation of \cite{ks} indicated that $\alpha$ was 
negative; we shall present a brief review of 
the argument below.

The starting point in \cite{ks} was that despite the background 
\scalim\ having a non-trivial dilaton profile, for purposes of 
computing the torus partition function all one needs to do is 
to replace the cigar by a cylinder of finite size, because the 
torus partition function is independent of the string coupling. 
The torus partition function being extensive, would be 
proportional to the volume of the cigar and also to the 
volume of the NS5-brane. The volume of the cigar is divergent, 
but can be regulated by introducing an ultra-violet cut-off
$ \phi \le \phi_{UV}$ (since this divergence is plainly associated with 
the semi-infinite linear dilaton CHS tube).
On the other hand presence of a regular horizon of the 
two-dimensional black hole (or in Euclidean signature, the tip of the cigar)
protects the CFT from entering the strong coupling region 
and also provides an IR cut-off. So forgetting about the dilaton 
profile we can replace the cigar by a cylinder of length $L_{\phi}$, 
and radius $\beta_H$, where

\eqn{length}{
L_{\phi} = \phi_{UV} - \phi(\sigma = 0)  = \phi_{UV} - { 1 \over 2} 
\log{\mu  \over N} = - { 1 \over 2} \log E + const. 
}

\noindent
So one is instructed therefore to compute the free string theory
partition function on 
\eqn{bgpf}{
{\bf R}_{\phi} \times {\bf S}^1 \times {\bf R}^5 \times SU(2)_N
}
\noindent
In the large $N$ limit one could replace the three-sphere by 
${\bf R}^3$, since the radius of the ${\bf S}^3$ scales like 
$\sqrt{N}$. So apart from a factor of $L_{\phi}$ which 
arises because of the extensivity of the partition function, 
the rest of the calculation amounts to computing free string theory 
partition function on ${\bf R}^9 \times {\bf S}^1$, which is 
basically the same computation as in \cite{aw}. 
The basic result obtained in \cite{ks} is that 

\eqn{pfn}{
-\beta F = Z_{torus} = {\beta V_5 L_{\phi} \over 4} \gamma
}

\noindent
where, $\gamma$ is a positive definite number that can be found in 
\cite{ks}; it is basically the free string partition function 
evaluated at the Hagedorn temperature of the LST \temp.
Assuming that the density of 
states is given as in \dos\ we can compute the free energy. It is 
of the same form as in \pfn, giving, 

\eqn{aldet}{
\alpha + 1 = - {\beta V_5 \over 4} \gamma
}

\noindent
Where we have made use of the fact that $L_{\phi} \sim - \log E$.
For an explicit expression the 
reader is referred to the original paper \cite{ks}. 

\section{A Question of Interpretation} 
The string theory computation shows that the thermal ensemble has 
an instability at one loop. Kutasov and Sahakyan \cite{ks},
interpreted this to  be indicative of a stringy mode becoming 
tachyonic at one loop. In the Lorentzian picture of the black hole, 
the thermal nature of the system appears to be a characteristic 
feature of the horizon. So a thermodynamic instability in the 
dual field theory could correspond to a deformation of the black hole 
horizon. This can be realized in the CFT if there were to exist 
a state which was potentially unstable to quantum corrections. 
One would in addition require that its wavefunction be
supported near the tip of the cigar, for it to 
correspond in the geometric language to a deformation of the horizon.
To summarize, one is looking for a state in the CFT which is 
massless at tree level, 
potentially becoming tachyonic when loop corrections are 
incorporated and localized near the tip of the cigar. 

A state that satisfies the above criteria was found in the analysis 
of DSLST \cite{gka,gkb}. It was argued that the two-point functions of 
operators ${\cal{O}}_m$, given in the $(-1,-1)$ picture as, 

\eqn{operator}{
{\cal{O}}_m = e^{-\varphi -\bar{\varphi} } V_{j;m,m} e^{i \vec{p}.\vec{x}}
}

\noindent
where, $\varphi,\bar{\varphi}$ are 
the bosonized ghosts, $\vec{p}$ is the spatial 
momentum along the longitudinal directions of the five-brane and 
$V_{j;m,m}$ is an observable in the $SL(2)/U(1)$ coset CFT of the cigar; 
have poles which correspond to light states. In particular there
exists a massless state which has unit winding number along the 
Euclidean time direction.  It was conjectured in \cite{ks}
that in the absence of spacetime supersymmetry this state is likely to 
acquire a tachyonic mass term from loop corrections and would therefore be 
a likely candidate for causing the instability.

The existence of a stringy instability in a 
smooth geometry wherein supergravity
is always valid, is extremely surprising. We would like to claim 
that this instability is in fact intrinsic to supergravity. To motivate 
this claim let us first of all understand how one can see the instability 
in supergravity. We have already mentioned that calculation of
supergravity one-loop 
partition function is rendered iffy by the usual problems of having to 
regulate divergences. But since string theory provides a natural 
regulator for supergravity, we shall view the zero mode contribution of the 
string theory partition function as the regulated 
supergravity partition function.

To check that the zero mode part of the string theory partition function has 
the same pathology, one starts from the sigma model with a non-trivial 
dilaton and separates out the zero mode piece of the dilaton from the 
fluctuation part and does the path integral over the zero mode. 
(this manipulation is standard in the context of 
two-dimensional Liouville theory, {\it cf.}, \cite{gtw, igor}).
This allows one to isolate the zero mode contribution in an explicit manner
and as we shall demonstrate reduce the rest of the problem to 
a familiar system.

Without loss of generality one can restrict oneself to the
cigar part of the geometry. What we have here is a   
Liouville field coupled to a single compact scalar;
\eqn{twodpf}{\eqalign{
Z_{torus}  &= \int [D\phi] [DX] e^{-S} \cr
S &= {1 \over 4 \pi } \int d^2 \sigma \sqrt{\hat{g}}
\left( \partial_{\mu} X \partial^{\mu} X + \partial_{\mu} \phi
\partial^{\mu} \phi - 4 \hat{R} \phi + 4 \Delta e^{-2 \phi} \right).
}}

\noindent
$\Delta$ here is a bare cosmological term and the hatted quantities 
denote background values. Splitting the $\phi$ integral into a 
zero mode and a fluctuation part; 
$ \phi = \phi_0 + \delta \phi$, and integrating out 
$\phi_0$ one obtains \cite{igor}, 

\eqn{pftd}{
Z_{torus} \sim \left({\Delta \over \pi }\right)^s
\Gamma(-s) \int [D\tilde{X}] [D\tilde{\phi}]  \; e^{-S_0} 
\left( \int d^2 \sigma \sqrt{\hat{g}} e^{- 2 \tilde{\phi}} \right)^s
} 

\noindent
where
\eqn{bareact}{
S_0 = \int d^2 \sigma \; \sqrt{\hat{g}} \left( 
\partial_{\mu} \tilde{X} \partial^{\mu} \tilde{X} + \partial_{\mu} 
\tilde{\phi} \partial^{\mu} \tilde{\phi} \right) 
}

\noindent
is the free action of corresponding to a compact scalar and a 
non-compact scalar. The quantity $s$ is basically related to the 
Euler characteristic of the world-sheet and also has contributions from the 
vertex operator insertions. For the particular case of the torus 
with no insertions one has $ s = 0 $, implying

\eqn{pffin}{
Z_{torus} = - { 1 \over 2} \log E \int_{\cal{F}} d^2 \tau 
\int D[\tilde{X}] [D\tilde{\phi}] e^{-S_0}.
}

\noindent
We obtained a factor proportional to $- { 1 \over 2 } \log E $
from the zero mode part (it is related to a properly regulated 
version of $\Delta^s$ in \pftd). What we have shown here is that the 
zero mode part of the partition function gives precisely the 
logarithmic dependence on energy with the right sign; the rest of 
the partition function is readily seen to be the free string theory 
contribution which has no pathologies for temperatures of order 
$T_H$. It is therefore tempting to argue for a supergravity origin of
the instability seen in the thermal ensemble.

\section{Supergravity Instability}
A well known example of classical instabilities 
in supergravity is the Gregory-Laflamme (GL) instability \cite{gla, glb}. 
The basic argument for a GL instability of a black p-brane is 
related to the fact that a black brane is entropically less favored 
when compared with a configuration comprising of an array of 
black holes, with the same amount of mass and charge per unit volume.
We might ask whether such an instability is likely to occur in the 
case of the LST background \scalim. At  first sight this appears unlikely. 
The GL instability was shown to exist only for branes 
that are far from extremal. 
In the original work \cite{glb} the numerical approximations were 
argued to be unreliable close to extremality. In the present 
case we are starting with non-extremal black 5-branes (we are 
in fact in the same set up as in \cite{glb}, since they are also 
consider NS-NS form fields), but the decoupling 
limit \decolim\ instructs us to take the non-extremality parameter 
$r_0$ to zero. Thus naively it would appear that one is beyond the 
regime of the GL instability that has been investigated and 
one might be forced to look elsewhere for the origins of the instability.

However, in a recent work by Reall \cite{reall}, it was argued that 
the GL instability persists for five-branes even in the near-extremal 
region (the result is also attributed to \cite{glc}). So one would 
have to be careful in claiming that the spacetime geometry \scalim, 
which is relevant for the decoupled LST, is GL stable.  
In fact we shall argue that in the limit $r_0 \rightarrow 0$, 
(note that since we also take $g_{str} \rightarrow 0$ we 
do not reach the extremal NS5-brane which is stable)
the GL instability does survive and is responsible for an 
unstable thermal ensemble. This would in line with the philosophy 
advocated in \cite{gma, gmb}. The argument made in these works 
was that a black brane would be classically
stable iff it were thermodynamically 
stable. In the present case given that we 
know of a thermodynamic instability in the dual field theory, 
we would like to associate it with a classical GL instability of the
supergravity background.

To make precise what exactly we mean by the geometry \scalim\ to be GL
unstable, let us briefly recall the analysis of \cite{glb}.
One starts with a classical black brane background, with a 
metric $g^{(0)}_{\mu \nu}$, dilaton $\phi^{(0)}$, and a p-form field strength
$F_{(p)}$. The black brane solutions are typically of the form 
$ { \bf R} \times { \bf R}^{8-p} \times { \bf S}^{p}$. Fluctuations 
about the classical background are considered, in particular one 
looks at fluctuations that carry some momentum along the longitudinal 
directions {\it i.e.},

\eqn{fluc}{\eqalign{
g_{\mu \nu }(r,t,x)  & = g^{(0)}_{\mu \nu}(r) \; + \; e^{i \omega t + i k.x} 
h_{\mu \nu}(r) \cr
\phi(r,t,x) & = \phi^{(0)}(r) \; + \; e^{i \omega t + i k.x} f(r) 
}}

\noindent
The instability is typically expected to occur in the s-wave sector 
of the sphere in question, for higher angular momenta modes would 
correspond to heavier particles of the KK spectrum on the sphere. 
Also fluctuations of the p-form field strength are set to zero
(see \cite{glb} for consistency of this ansatz at first order in 
perturbations). 
What the GL instability means is that there are solutions to the 
linearized equations about the black brane background, wherein 
the fluctuations are perfectly regular both at the horizon and in the 
asymptotic region, but have an imaginary frequency {\it i.e.}, the 
fluctuations grow exponentially in time as opposed to being 
oscillatory. Of course, this behavior is not present at all length scales.
In fact for small wavelengths there is no instability and the only 
regular solutions are those that are oscillatory in time. 
But above a critical wavelength corresponding to a critical 
longitudinal momentum $k_*$ (where $ k^2 = k_i.k^i$), there is an 
instability {\it i.e.}, the fluctuation equations admit 
regular solutions that are exponentially growing in time. At the 
critical momentum $k_*$ it is argued in \cite{reall} that there 
ought to be a time independent solution to the fluctuation equations, 
{\it i.e. } a zero frequency mode, which we shall refer to as 
the {\it threshold unstable mode}, borrowing Reall's terminology.

The analysis of \cite{reall} shows that when the fluctuation equations 
admit a constant time solution, the corresponding Euclidean action is 
negative definite. This  indicates a thermodynamic instability, for one can 
view the Euclidean action evaluated about the background to be the 
thermal partition function {\it a la } Gibbons-Hawking \cite{gibhawk}.
In particular the fact that the NS5-branes are GL unstable all the 
way down to extremality implies the existence of a threshold 
unstable mode for NS5-branes. We wish to claim that this threshold 
unstable mode survives the decoupling limit and is responsible for the 
instability of the thermal ensemble. The conjecture therefore 
can be stated as 

{\it In the decoupling limit of non-extremal NS5-branes, the 
threshold unstable mode \cite{reall} survives. This mode has zero frequency, 
but it potentially can acquire a tachyonic mass term at one-loop.
It is precisely this mode that is responsible for the 
thermodynamic instability seen in the partition function of \cite{ks}.}

At the present juncture we have no explicit proof for the existence of 
a threshold unstable mode. We hope to demonstrate the existence of such
in the near future. 

\section*{Acknowledgments}
It is a great pleasure to thank O.~Aharony, M.~Berkooz,
S.~Gubser, G.~Horowitz, V.~Hubeny, I.~Klebanov, 
K.~Narayan, I.~Mitra and especially 
E.~Witten for discussions. In addition I would
like to thank Caltech for hospitality, where this work was initiated.
This work was supported in part by NSF grant PHY-980248.

\end{document}